# Optical Encryption with Jigsaw Transform, Using MATLAB

Leidy Marcela Giraldo Castaño          Edward Yesid Villegas Pulgarín
lgiral25@eafit.edu.co                              evillega@eafit.edu.co
Logic and Computation Group
Program of Physical Engineering
EAFIT University, Medellín, Colombia

**Abstract -** *This article will describe an optical encryption technical of images which it is proposed in an analogical and digital way. The development of the technical to a digital level, it is made to implementing algorithms (routines) in MATLAB.*

*We will propose a functional diagram to the described analogical development from which designated the optical systems associated with each functional block.*

*Level of security that the jigsaw algorithms provide applied on an image, which has been decomposed into its bit-planes, is significantly better if they are applied on an image that has not been previously decomposed.*

## 1   Introduction

Nowadays, one of the issues that driving the development of applications and improving existing methods in the fields of optics, optoelectronics and others, is the protection of information in a data transfer. That is achieved through encryption of information based usually in mathematical definitions, which can lead its material implementation.

Thus, a lot of information encryption methods have emerged in a several fields, using mathematical tools such as The Fractional Fourier Transform (FRT), Hadamark Transform, Walsh Codes, Orthogonal Codes and Jigsaw Transform (JST) or even combinations of them.

The security offered by these methods is related to the number of transformation implemented, the number of parameters associated with those and the behavior of each transform.

## 2   Problem

If we want to make an optical encryption for an image using the JST, the first step is to define a JST from its proporties. The JST rearranges the blocks of a complex image by randomly way. The JST is an unitary transform and we can find an inverse transform for it.
There are some kind of JST based on the number of dimension (e.g. 2D JST and 3D JST [Figure 1]) used to rearrange the blocks of the image and the particular routine to apply the transform on an image.

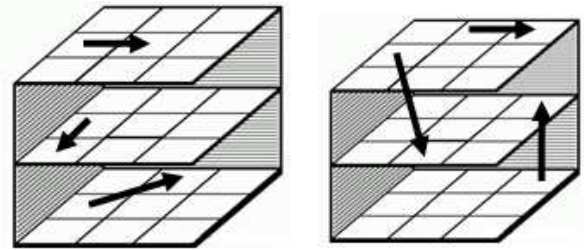

Figure 1.   2D-JST          3D-JST

The *Figure 1 shows* a scheme for a decomposed image in its bit–planes and the action of the JST on the image.

The Bit-Planes of an image are a complex of layers, where each layer includes the intensity of an specific range of the gray scale.

## 3   Method

The procedure begins when the image that you want to encrypt is converted in gray scale, after that, the image is decomposed in bit – planes and then is applied the JST on each one of them with a different parameter. Finally, the jigsawed Bit-planes are multiplexed to obtain an encrypted image. The procedure in block diagrams is showed in the *Figure 2.*

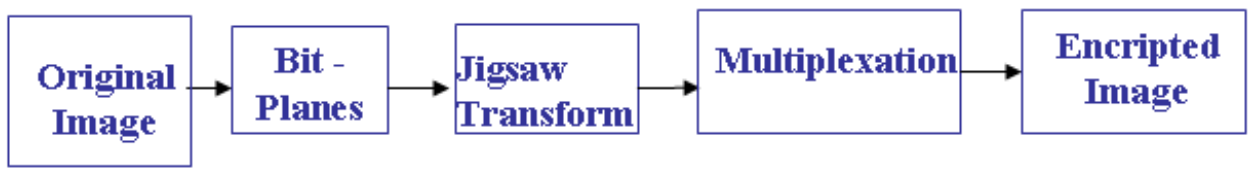

*Figure 2.* Blocks Diagram for a digital encryption using JST

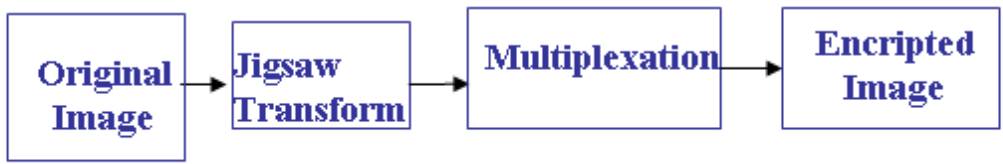

*Figure 3.* Blocks Diagram for a digital encryption using JST without separation in Bit-Planes

If the image is not decomposed in its bit-planes, the method only requires the direct application of the JST, that is showed in *Figure 3*.

The JST does not necessarily act pixel-pixel, it could act on blocks with nxm pixels that they form the image.

We had to develop routines in MATLAB 7.4 to implement the decomposition in bit-planes for an image that we wanted to encrypt, to apply the JST and to multiplex the jigsawed bit-planes.

## 4  Results

The original image that we want encrypt is showed in the *Figure 4*.

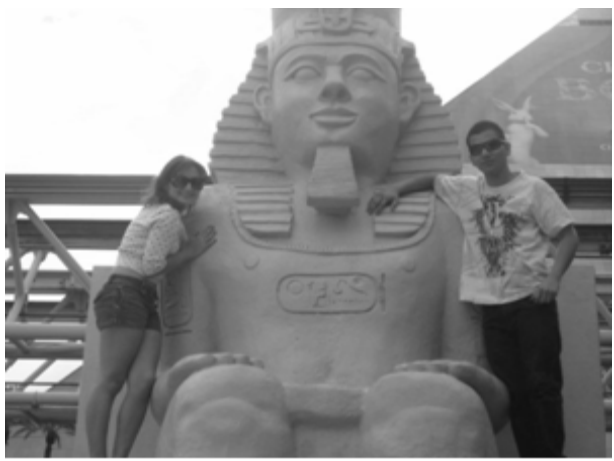

*Figure 4.* Gray scale Image

The results obtained implementing the separation in Bit-Planes is showed in *Figure 5*.

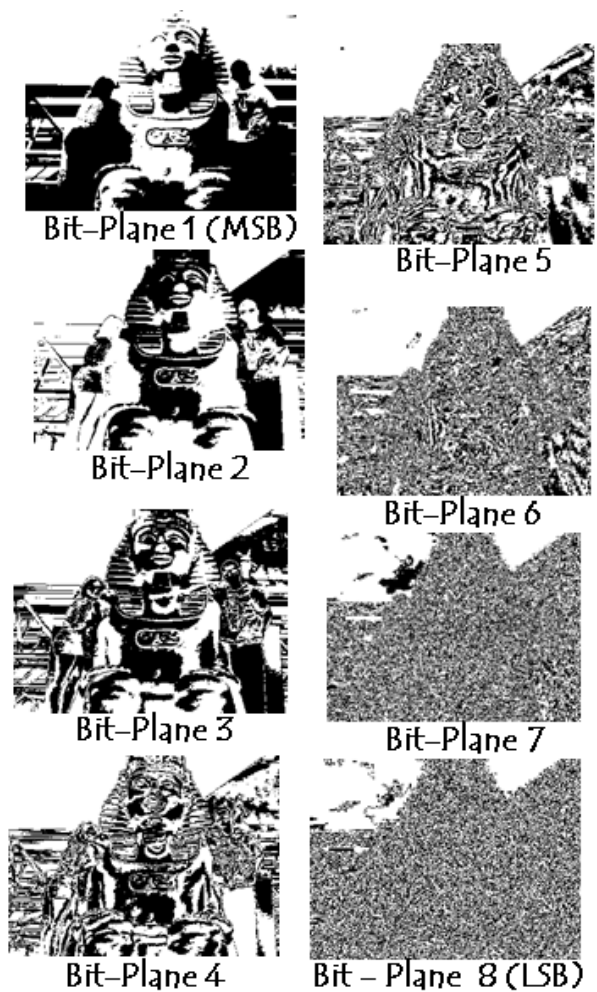

*Figure 5.* Bit - Planes

When we apply the JST with a different parameter on each one of the Bit- planes, we obtain the result showed in the *Figure 6*.

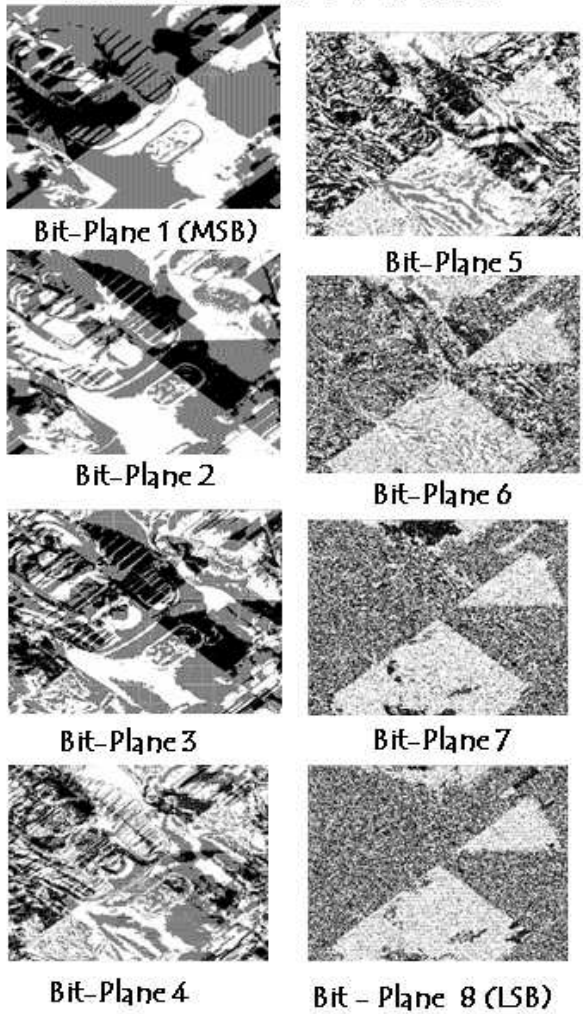

Figure 6. Jigsawed Bit-planes

After the multiplexing procedure, the encrypted image is obtained. That image is showed in the *Figure 7 and Figure 8*.

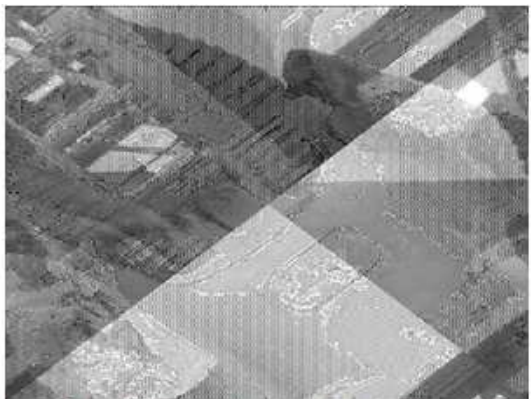

Figure 7. Multiplexed Image after to apply 2D-JST on each one of the Bit-Planes = Encrypted Image

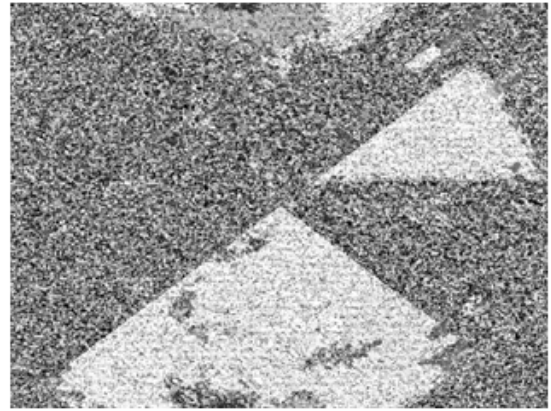

Figure 8 Multiplexed Image after to apply 3D-JST on each one of the Bit-Planes = Encrypted Image

If we decided to encrypt the image without a Bit-Plane separation and we apply the JST on the image the result obtained is showed in *Figure 9*.

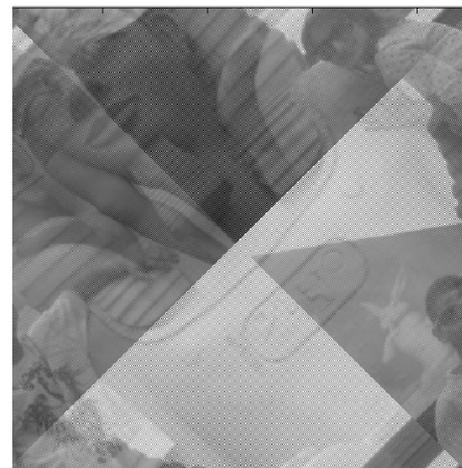

Figure 9. Encrypted image applying JST once.

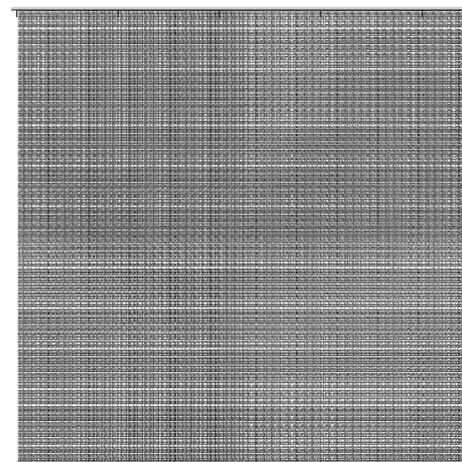

Figure 10. Encrypted image applying JST 6 times with the same parameter.

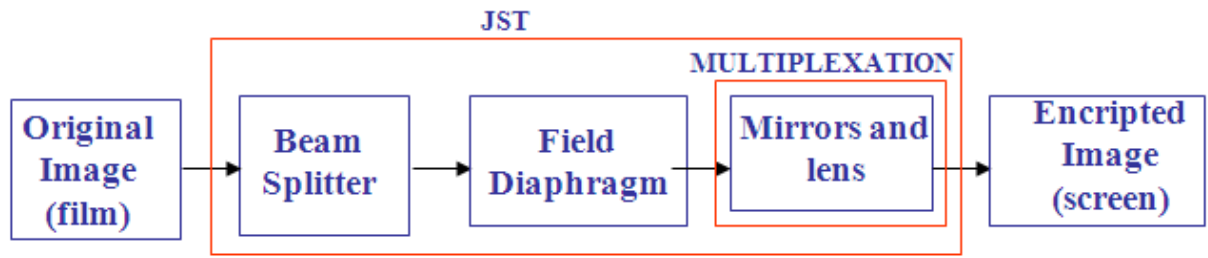

Figure 11. Analogical Way to encrypt an image without separation in Bit-planes

If we want to improve the encryption with this method, we can apply many times the JST with the same parameter or with another (See *Figure 10*).

We also propose a method to implement an analogical procedure to encrypt an image. The scheme in the *Figure 11 shows* an analogical way for that.

## 5 Conclusions

The security offered by these methods is related to the times that we apply the JST on the image, the number of parameters associated with the JST, the kind of the JST and the number of bit planes in which we decomposed the image.

Level of security that the jigsaw algorithms provide applied on an image, which has been decomposed into its bit-planes, is significantly better if they are applied on an image that has not been previously decomposed.

The level of security on the key of jigsaw algorithms based on Bit-Planes can improve significantly using the third dimension for the randomization of data blocks, also is possible to add random phase codes and / or Fractional Fourier Transform (FRT) to increase its security.